\documentclass[pra,preprint,tightenlines,showpacs,twocolumn,10pt]{revtex4} 
\usepackage{epsfig} 

\begin{document}  
    
\title{ Dynamics of correlated transfer-ionization in collisions with a fast highly charged ion }
 
\author{ A.B.Voitkiv }   
\affiliation{ Max-Planck-Institut f\"ur Kernphysik, 
Saupfercheckweg 1, D-69117 Heidelberg, Germany } 

%\date{\today} 

\begin{abstract} 
Transfer-ionization in fast collisions between 
a bare ion and an atom, in which one of 
the atomic electrons is captured by the ion 
whereas another one is emitted, 
crucially depends on dynamic electron-electron correlations.
We show that in collisions with a highly charged ion 
a strong field of the ion has a very profound effect 
on the correlated channels of transfer-ionization.
In particular, this field weakens     
electron emission into the direction opposite 
to the motion of the ion 
and strongly suppresses the emission 
perpendicular to this motion. Instead, 
electron emission is redirected into those parts 
of the momentum space which are  
very weakly populated in fast collisions 
with low charged ions. 
 
\end{abstract} 


\maketitle 

%% body of paper begins here 

%\newpage 

\section{Introduction} 

Electron-electron interaction  
is responsible for very many phenomena studied by 
the different fields of physics ranging from astrophysics  
to biophysics. Amongst them atomic physics 
and its part -- physics of ion-atom collisions -- often deal  
with most basic and clear manifestations of this interaction. 

Atomic excitation and ionization \cite{eich}-\cite{croth}, 
projectile-electron excitation and loss 
\cite{mcg}-\cite{sdr}, \cite{abv-buch},  
electron transfer (capture) \cite{eich}, \cite{croth}  
and pair production \cite{eich}, \cite{croth}-\cite{abv-buch}    
belong to the elementary reactions occurring when 
a projectile-ion collides with a target-atom. 
A combination of these reactions  
in a single-collision event is also possible 
and in such a case the electron-electron interaction 
during the collision when the external field 
is rapidly changing (dynamic electron correlations)  
is often crucial. 
 
In particular, mutual ionization in which 
both a target-atom and a (partially stripped)  
projectile-ion eject electrons, 
and transfer-ionization in which one of the atomic electrons 
is captured by a projectile-ion whereas 
another one is emitted, represent 
processes where dynamic electron correlations     
play a crucial role \cite{mcg}-\cite{sdr}, \cite{abv-buch}, 
\cite{mut-ioniz}, \cite{thomas}-\cite{ich-EE}. 

Transfer-ionization in fast collisions of low charged ions 
(mainly protons) with helium is attracting much attention 
\cite{mergel}-\cite{schoeffler}. This process can be analyzed in terms 
of different reaction mechanisms which are characterized 
by distinct features in the electron emission pattern. 
Depending on whether the electron-electron 
interaction plays in them a crucial role, 
these mechanisms can be termed "correlated" 
or "uncorrelated". 

Uncorrelated mechanisms are  
independent transfer-ionization (ITI) 
and capture--shake-off (C-SO). 
In the ITI electron capture and emission occur 
due to "independent" interactions of the 
projectile with two target electrons.    
According to the C-SO,  
a fast transfer of one electron from 
the atom to the ion leads to a "sudden" change  
of the atomic potential for another electron 
that leads to its emission. 
Both ITI and C-SO result in emission 
of low-energy electrons.    

The correlated mechanisms include  
electron-electron Thomas (EET) and 
electron-electron Auger (EEA).  
Within the EET transfer-ionization proceeds 
\cite{thomas}, \cite{briggs}-\cite{mcguire-tolmanov} 
via a binary collision of the projectile 
with one of atomic electrons and a consequent rescattering 
of this electron on another atomic electron. 
After these two collisions one 
of the electrons moves together with the projectile 
(that makes capture probable) 
while the other is emitted perpendicular to 
the projectile motion.  

The functioning of the EEA mechanism is based 
on the fact that merely the presence of  
the projectile makes the target unstable 
with respect to a kind of Auger decay.    
Indeed, viewing the collision in the rest 
frame of the projectile we can see that 
one of the electrons, belonging initially 
to a bound configuration of moving 
particles constituting the atom, can make 
a transition into a bound state of the ion 
by transferring the energy excess 
to another atomic electron which, 
as a result of this, is emitted from 
the atom in the direction 
of the atomic motion \cite{ich-EE}, \cite{we-EE}. 
In the rest frame of the atom this electron 
finally moves in the direction opposite 
to the projectile velocity 
\cite{ich-EE}, \cite{we-EE}, \cite{voit-2012}. 

The mechanisms, briefly discussed above, were proposed for  
describing transfer-ionization in collisions between 
a light atom and a low charged ion  
moving with a velocity $v$, which is much higher 
than the typical orbiting velocities of the electron(s)  
in their initial and final bound states: 
$v \gg Z_a |e|/\hbar  $ and $v \gg Z_i |e|/\hbar$, 
where $Z_a$ and $ Z_i$ are the charges of the nuclei 
of the atom and ion, respectively. 

What, however, can one say about transfer-ionization 
in fast collisions with highly charged ions (HCIs) when 
the charge $Z_i$ of the ion is so large that $Z_i \sim \hbar v/|e| $ ?  
One can expect that in such collisions, which are characterized by 
very strong fields generated by the HCI, not only cross sections 
for transfer-ionization would be much larger than in collisions with 
equivelocity low charged ions but also new interesting features 
could arise in this process.

Therefore, in this article we explore 
transfer-ionization in fast collisions with HCIs. It will be seen 
below that a strong field of HCI has a dramatic effect 
on the correlated channels of transfer ionization: 
it weakens the EEA, eliminates the EET and leads to qualitatively new structures in the emission spectrum.       
Atomic units ($\hbar = m_e = |e| =1 $) are used throughout  
except where the otherwise stated.   
 
\section{General consideration }  

\subsection{Correlated transfer-ionization } 
 
We are mainly interested in the correlated 
transfer-ionization and begin with its treatment. 
This treatment will be semiclassical in which only the  
electrons are described quantum mechanically whereas 
the heavy particles (the nuclei of the ion and atom) are 
considered classically. In fast collisions the trajectories of 
the heavy particles are practically straight-line. 
It is convenient to make the basic consideration of the correlated 
transfer-ionization using the rest frame of the ion and 
to take its position as the origin.  

According to scattering theory the  
exact (semiclassical) transition amplitude 
can be written as  
\begin{eqnarray} 
a_{fi} = -i \int_{-\infty}^{+\infty} dt % 
\langle \psi_f(t) |\hat{W}(t) |\Psi_i^{(+)}(t)\rangle.      
\label{e1} 
\end{eqnarray}  
Here $\Psi_i^{(+)}$ is an exact solution of 
the time-dependent Schr\"odinger equation 
with the full Hamiltonian $\hat{H}$ which describes  
two electrons moving in the external field 
of the nuclei and interacting with each other, 
$\psi_f $ denotes the final state of the two electrons 
and $\hat{W}$ is that part of $\hat{H}$ 
which is not included into the wave equation 
for $\psi_f$. Since the contribution to transfer-ionization 
from collisions, in which electrons change spin,  
is negligible we shall disregard 
the spin parts of the states $\Psi_i$ and $\psi_f$.  

In the correlated transfer-ionization the velocities of 
the electrons with respect to the nucleus of the atom  
in the final state are of the order of $v$ \cite{voit-2012}. 
Besides, in this state the relative velocity 
of the electrons is also of the same order. 
Therefore, when this process occurs in fast collisions with HCIs, 
for which one has $Z_i \sim v$ but $ \max\{Z_a,1\} \ll v$,   
the motion of both electrons in the final state $\psi_f$ 
is driven by the field of the HCI whereas the interactions 
of the electrons with the atomic nucleus 
and with each other can be neglected. 
Thus, we have   
\begin{eqnarray} 
\psi_f(t)&=& \frac{1}{\sqrt{2}}  
\left( \chi_b({\bf r}_1) \chi_{\bf p}({\bf r}_2) 
\pm \chi_b({\bf r}_2) \chi_{\bf p}({\bf r}_1) \right)     
\nonumber \\ 
&& \times \exp(- i (\varepsilon_f + p^2/2)t),  
\label{e2} 
\end{eqnarray}  
where ${\bf r}_1$ and ${\bf r}_2 $ are the coordinates 
of the electrons, $\chi_b$ is the bound state of 
an electron captured by the HCI with an energy $\varepsilon_f$ and 
$\chi_{\bf p}$ is the state of emitted electron 
which moves in the HCI's field and has  
asymptotically a momentum ${\bf p}$.  

When the state $\psi_f$ is taken in the form (\ref{e2}) 
the perturbation $\hat{W}$ in Eq.(\ref{e1}) is equal to 
$\hat{W}_{1a} + \hat{W}_{2a} + \hat{W}_{12} $,  
where $\hat{W}_{ja}$ ($j=1,2$) is the interaction   
between the $j$-th electron and the nucleus of the atom 
and $\hat{W}_{12}$ is the electron-electron interaction.  
Then we obtain 
\begin{eqnarray} 
a_{fi} &=& -i \int_{-\infty}^{+\infty} dt % 
\langle \psi_f(t) | \hat{W}_{1a} + \hat{W}_{2a} |\Psi_i^{(+)}(t)\rangle 
\nonumber \\ 
&& - i \int_{-\infty}^{+\infty} dt 
\langle \psi_f(t) | \hat{W}_{12} |\Psi_i^{(+)}(t)\rangle,       
\label{e3} 
\end{eqnarray}  
It is the last term on the right-hand side of Eq.(\ref{e3}) 
which is relevant for the correlated transfer-ionization.   

In order to find a suitable approximation 
for $\Psi_i^{(+)}(t)$ let us note the following.   
In the process of correlated transfer-ionization occurring 
in very fast collisions ($v \gg Z_a$) both electrons undergo transitions 
in which the change in their momenta is much larger than their 
typical momenta in the initial atomic state. Because of that, 
even if the projectile would have a low charge, the nucleus of the atom 
would be merely a spectator during this process \cite{voit-2012}. 
Under such circumstances the so called impulse approximation, 
in which the role of the atomic nucleus is just to produce 
the momentum distribution (and binding energy) of the electrons 
in the initial state, can be used to treat transfer-ionization. 

In our case, in which transfer-ionization occurs in 
very asymmetric collisions where the charge of the HCI 
is much higher than the charge of the atomic nucleus, 
the impulse approximation is even more appropriate because 
it enables one to fully account for the influence 
of the strong field of the HCI on {\it two electrons}. 

The initial (undistorted) atomic two-electron state is given by 
\begin{eqnarray} 
\Psi_i^0(t)&=&\varphi_a({\bf r}_1-{\bf R}_a(t),{\bf r}_2-{\bf R}_a(t)) 
\exp(i{\bf v}_a \cdot ({\bf r}_1 + {\bf r}_2)) 
\nonumber \\ 
&& \times \exp(-i v_a^2 t ) \exp(-i\epsilon_a t),       
\label{e4} 
\end{eqnarray}  
where ${\bf R}_a(t)= {\bf b} + {\bf v}_a t$
is a straight-line classical trajectory of the atomic nucleus  
moving with a velocity ${\bf v}_a$ and $\varphi_a$ is 
the initial atomic state with an energy $\epsilon_a$ 
(as viewed in the rest frame of the atom). 

Now we perform the three-dimensional Fourier transformation 
of the state $\varphi_a({\bf r}_1-{\bf R}_a(t),{\bf r}_2-{\bf R}_a(t))$, 
insert the obtained result into Eq.(\ref{e4}) and replace 
the plane-wave factors in the integrand by the corresponding 
Coulomb waves in the field of the HCI. In such a way we approximate 
the state $\Psi_i^{(+)}(t)$ in Eq.(\ref{e3}) by  % $\Psi_i^{IA}(t)$ 
\begin{eqnarray} 
\Psi_i^{IA}(t)&=& \frac{\exp(-i (v_a^2 + \epsilon_a) t)}{(2 \pi)^3} 
\int d^3 \mbox{\boldmath$\kappa$}_1  
\int d^3 \mbox{\boldmath$\kappa$}_2 \, \,   
\phi_a(\mbox{\boldmath$\kappa$}_1,\mbox{\boldmath$\kappa$}_2)  
\nonumber \\ 
&& \times \exp(-i(\mbox{\boldmath$\kappa$}_1 + \mbox{\boldmath$\kappa$}_2) 
\cdot {\bf R}_a(t))
\chi_{{\bf p}_1}({\bf r}_1) 
\chi_{ {\bf p}_2 }({\bf r}_2).     
\label{e5} 
\end{eqnarray}  
Here $ \phi_a $ is the Fourier transform of the state $\varphi_a$,  
${\bf p}_1 = {\bf v}_a+\mbox{\boldmath$\kappa$}_1$ 
and ${\bf p}_2 = {\bf v}_a+\mbox{\boldmath$\kappa$}_2$ 
are the initial momenta of the electrons with respect to the HCI 
and $\chi_{{\bf p}_j}({\bf r}_j) $ ($j=1,2$) are 
Coulomb wave functions of electrons which move 
in the field of the HCI being incident on it with  
asymptotic momenta ${\bf p}_j$.   

Using Eqs.(\ref{e2})-(\ref{e3}) and (\ref{e5}) and assuming 
that the space part of the initial atomic state is symmetric 
with respect to the interchange of the electrons 
(like it is the case for the ground state of helium)   
one can show, after some algebra, 
that in the projectile frame 
the cross section for transfer-ionization 
differential in the momentum ${\bf p}$ of the emitted electron 
is given by  
\begin{eqnarray} 
\frac{ d \sigma }{ d^3 {\bf p} }& = &\frac{1}{16 \pi^2 v^2} 
\int d^2 {\bf q}_{\perp}  
\nonumber \\ 
&& \left| \int d^3 \mbox{\boldmath$\kappa$} 
\phi_a\left(\frac{{\bf q} + \mbox{\boldmath$\kappa$} }{2}, 
\frac{{\bf q} - \mbox{\boldmath$\kappa$} }{2}\right) \, W_{fi} 
\right|^2.        
\label{e6} 
\end{eqnarray}  
In this expression   
\begin{eqnarray} 
{\bf q} = \left({\bf q}_{\perp}, 
\frac{ \varepsilon_f + p^2/2 - v_a^2 - \epsilon_i }{ v_a }\right)     
\label{e7}  
\end{eqnarray}  
is the momentum transfer in the collision with 
${\bf q}_{\perp}$ being its transferse part 
(${\bf q}_{\perp} \cdot {\bf v}_a =0$) and  
\begin{eqnarray} 
W_{fi} = \langle    
\chi_b({\bf r}_1) \chi_{\bf p}({\bf r}_2)  
\left| \frac{1}{r_{12}} \right| 
\chi_{{\bf p}_1}({\bf r}_1) \chi_{{\bf p}_2}({\bf r}_2) \rangle,  
\label{e8} 
\end{eqnarray}  
where ${\bf p}_1 = {\bf v}_a + ({\bf q} + \mbox{\boldmath$\kappa$})/2 $ 
and ${\bf p}_2 = {\bf v}_a + ({\bf q} - \mbox{\boldmath$\kappa$})/2 $. 
  
Expression (\ref{e6}) can be drastically simplified by using 
the fact that the Fourier transform 
$\phi_a\left(({\bf q} + \mbox{\boldmath$\kappa$})/2, 
({\bf q} - \mbox{\boldmath$\kappa$})/2\right) $ becomes very small 
when the absolute values of $ {\bf q} \pm \mbox{\boldmath$\kappa$} $
substantially exceed the typical electron velocities inside the atom.  
Since we consider collision velocities, which 
are much higher than the latter ones, 
we may set in Eq.(\ref{e8}) ${\bf p}_1 = {\bf p}_2 = {\bf v}_a$ 
and take $|W_{fi}|^2$ in Eq.(\ref{e6}) out of the integrals.

\subsection{Uncorrelated transfer-ionization}  
  
Let us now say few words about the treatment of 
the uncorrelated channels of transfer-ionization. 
Following \cite{ich-EE}, \cite{we-EE} and \cite{voit-2012} 
we shall present the amplitude for this process 
as the product of single-electron transition 
amplitudes for capture and ionization which are 
obtained using three-body models: continuum-distorted-wave 
(for capture) and continuum-distorted-wave-eikonal-initial-state 
(for ionization) \cite{croth}. 

Because of a high charge of the projectile 
the C-SO, compared to the ITI, contributes negligibly.  
Since the emission produced by these channels 
is localized in the same part of the momentum space 
the C-SO can safely be neglected. 

\section{ Results and discussion } 
  
\begin{figure}[t] 
\vspace{-0.45cm}
\begin{center}
\includegraphics[width=0.5\textwidth]{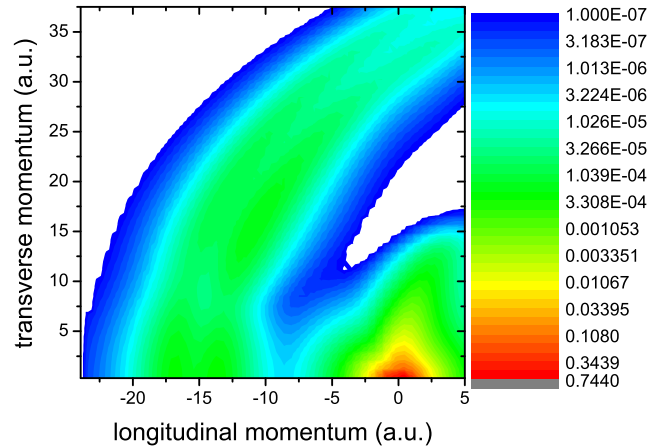}
\end{center}
\vspace{-0.9cm}  
\caption{ \footnotesize{ Momentum spectrum (in b/(a.u.)$^3$) 
of electrons emitted in the reactions     
$22.5$ MeV Ca$^{20+}$ + He(1s$^2$) $\to$ $\sum_{n=1}^2$ Ca$^{19+}$(n) + 
He$^{2+}$ + e$^-$ collisions ($v=30$ a.u.). }}  
\label{figure1} 
\end{figure}

\begin{figure}[t] 
\vspace{-0.45cm}
\begin{center}
\includegraphics[width=0.5\textwidth]{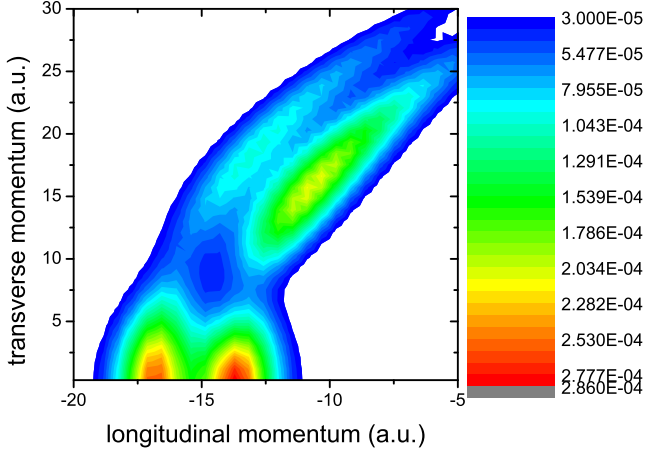}
\end{center}
\vspace{-0.9cm} 
\caption{ \footnotesize{ Same collision system 
as in figure \ref{figure1} but only 
the emission via the correlated channels is shown. }}  
\label{figure2} 
\end{figure}

In figures \ref{figure1}-\ref{figure2} 
we present results \cite{details} for 
the momentum spectrum of electrons  
emitted in transfer-ionization in collisions 
of $22.5$ MeV/u Ca$^{20+}$ projectiles ($v=30$ a.u.) 
with helium atoms when capture 
occurs into the $K$ and $L$ shells. The spectrum is given 
in the rest frame of the target 
($=$ the laboratory frame) in which the projectile moves 
with a velocity ${\bf v}$ and is represented 
by the doubly differential cross section 
\begin{eqnarray} 
\frac {d^2 \sigma }{ k_{tr}  dk_{lg} dk_{tr} } =  
\int_0^{2 \pi} d\varphi_k \,  
\int d^2 {\bf q}_{\perp} \left| S_{fi}({\bf q}_{\perp}) \right|^2,        
\label{e9} 
\end{eqnarray}  
where $k_{lg} = {\bf k} \cdot {\bf v} /v $ and 
${\bf k}_{tr} = {\bf k} - k_{lg} {\bf v} /v  $ 
are the longitudinal and transverse parts, 
respectively, of the momentum ${\bf k}$  
of the emitted electron in the laboratory frame 
and $k_{tr}=|{\bf k}_{tr}|$. 
The integration in (\ref{e9}) runs over the transverse part of 
the momentum transfer % ${\bf q}$ 
and the azimuthal angle 
$\varphi_k$ of the emitted electron. 

In figure \ref{figure1} the maximum at small 
momenta has its origin in the ITI whereas  
the maxima at much larger $k$ appear due 
to the correlated channels.    
The maximum at small momenta yields 
a very important contribution to 
the total cross section. However, its structure 
is similar to that in fast collisions with 
low charged projectiles, studied in  
\cite{voit-2012}, and below will not be considered.  

The maxima at large $k$ deserve more attention. 
In order to see better their structure,  
in figure \ref{figure2} only the high-momentum part 
of the emission, which is produced solely 
via the correlated channels, is shown. 

In the rest frame of the ion the approximate 
energy balance for transfer-ionization is very simple: 
$  v^2 + \epsilon_a \approx \varepsilon_f + p^2/2 $, 
where $\varepsilon_f = - Z_i^2/2n^2$ with $n$ being the principal 
quantum number of the bound state of the captured electron. 
Since $ {\bf k}_{tr} = {\bf p}_{tr} $ and $k_{lg} = v - p_{lg}$, where 
$ {\bf p}_{tr} $ and $p_{lg}$ are respectively 
the transverse and longitudinal parts of ${\bf p}$, 
in the target frame the emission is concentrated 
on ridges located along rings with radii  
$R_n= \sqrt{2v^2 + \epsilon_a + Z_i^2/2n^2 }$ 
centered at ($k_{lg}=v$, $k_{tr}=0$). 
The ridge with higher energy appears 
due to transfer-ionization with electron 
capture into the ground state of the projectile 
while the other ridge originates from 
capture into the projectile's $L$-shell. 
It is seen in the figure that each of these ridges  
has two distinct maxima: one centered at an emission 
angle of $\vartheta_k = 180^0$ and the other at 
$\vartheta_k \approx 125^0$ (the angle $\vartheta_k$ 
is counted from the direction of the projectile motion).  

\begin{figure}[t] 
\vspace{-0.45cm}
\begin{center}
\includegraphics[width=0.5\textwidth]{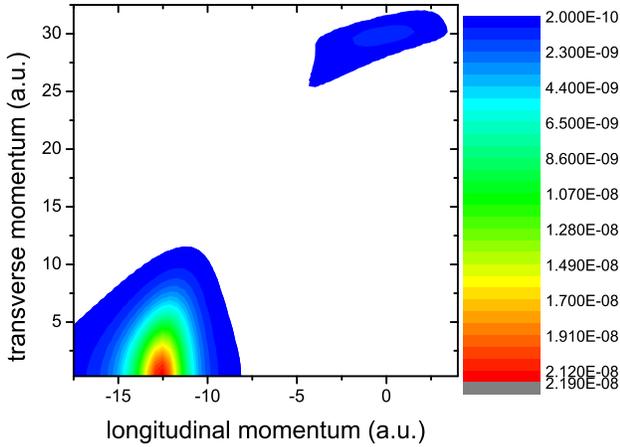}
\end{center}
\vspace{-0.9cm} 
\caption{ \footnotesize{ Momentum spectrum (in b/(a.u.)$^3$) 
of electrons emitted via the correlated mechanisms 
in the reaction $22.5$ MeV p$^{+}$ + He(1s$^2$) $\to$ H(1s) + 
He$^{2+}$ + e$^-$. }}  
\label{figure3} 
\end{figure}

The shape of the correlated part of the spectrum 
in case of collisions with HCIs is to be compared with that 
in collisions with low charged ions. 
The latter is displayed in figure \ref{figure3} 
for $22.5$ MeV p$^{+}$ + He(1s$^2$) collisions. 
Since at $Z_i \ll v$ transfer-ionization 
is strongly dominated by capture 
into the ground state this spectrum \cite{plane-waves} 
is concentrated on a single ridge. 
Compared to the corresponding ridge 
in \ref{figure2} the ridge in figure \ref{figure3} 
is shifted to lower $k$ because of 
the much smaller binding energy of the captured electron. 
It consists of two distinct parts:     
the maximum at $\vartheta_k = 180^0$  
is caused by the EEA whereas the maximum 
at $\vartheta_k \approx 90^0$ is the signature 
of the EET.  
 
Comparing the spectra in figures \ref{figure2} and \ref{figure3}, 
one can attribute the maxima at $\vartheta_k = 180^0$  
in figure \ref{figure2} as arising due to the EEA. 
However, in the strong-field regime 
there is no maximum at $\vartheta_k \approx 90^0$ 
which is characteristic of the EET channel 
in collisions with fast low-charged ions. 
Instead, a new maximum appears on each ridge 
at $\vartheta_k \approx 125^0$ 
which is absent when the projectile has a low charge. 
This maximum is rather intense and its contribution 
to the total cross section is comparable to (or even exceeds)  
that of the EEA maximum.    

\begin{figure}[t] 
\vspace{-0.45cm}
\begin{center}
\includegraphics[width=0.55\textwidth]{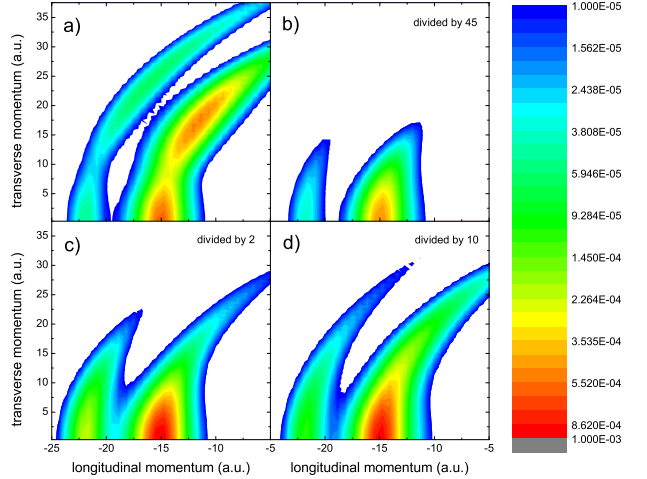}
\end{center}
\vspace{-0.5cm} 
\caption{ \footnotesize{ The momentum spectra for 
the reactions 
$22.5$ MeV/u Zn$^{30+}$ + He(1s$^2$) 
$\to$ $\sum_{n=1}^2 $ Zn$^{29+}$(n) + 
He$^{2+}$ + e$^-$. }}  
\label{figure4} 
\end{figure}

In order to get more ideas about the correlated 
transfer-ionization in the strong-field regime, in figure 
\ref{figure4} we present the emission spectrum 
at even a stronger field when 
$22.5$ MeV/u Zn$^{30+}$ projectiles collide with helium. 
As before, we consider capture into the $K$ and $L$ shells only.   
From figure \ref{figure4}a it can be inferred 
that the structure of the spectrum is similar to that 
in collisions with Ca$^{20+}$ ions: it has two ridges 
and each of them has two pronounced maxima centered at  
$\vartheta_k \approx 125^0$ and $ 180^0 $. 
However, the relative intensity 
of these maxima is now different 
with the maxima at $\vartheta_k \approx 125^0$ 
becoming noticeably more populated 
compared to those at $ \vartheta_k = 180^0$. 
Besides, the relative importance of the capture into 
the ground state descreases (as expected).  

Figure \ref{figure4} shows also three more results. 
The spectrum presented in \ref{figure4}b was obtained 
by approximating all the states $\chi_{{\bf v}_a}({\bf r}_1)$, 
$\chi_{{\bf v}_a}({\bf r}_2)$ and $\chi_{{\bf p}}({\bf r}_2)$
by plane waves, i.e. assuming that both electrons in the initial 
channel as well as the emitted electron do not feel 
the HCI's field.  

Figure \ref{figure4}c displays results calculated 
when the states $\chi_{{\bf v}_a}({\bf r}_2)$ 
and $\chi_{{\bf p}}({\bf r}_2)$ are modelled by plane waves 
whereas the state $\chi_{{\bf v}_a}({\bf r}_1)$ is the coulomb one. 
Thus, in this case we neglect the action of the HCI's 
field only on that electron, which is not captured.  

Finally, figure \ref{figure4}d shows the spectrum 
obtained if the state $\chi_{{\bf v}_a}({\bf r}_1)$ is taken 
as a plane wave but the states $\chi_{{\bf v}_a}({\bf r}_2)$ 
and $\chi_{{\bf p}}({\bf r}_2)$ are coulomb waves. 

When the action of the HCI's field  
is neglected for both electrons 
(of course, except in the final bound state $\chi_b$) 
the calculated spectrum has very pronounced 
maxima at $ \vartheta_k = 180^0$ and their 
intensity rapidly decrease when 
the transverse component $k_{tr}$ 
of the electron momentum increases 
(see figure \ref{figure4}b). 

If the field of the HCI 
is neglected only for that electron, 
which is finally emitted,  
the maxima at $ \vartheta_k = 180^0$ become 
less pronounced but new maxima do not yet 
appear although the calculated spectrum 
has more extension in the direction of larger $k_{tr}$ 
(see figure \ref{figure4}c). 
If we neglect the action 
of the HCI's field on that electron 
which is finally captured but take 
into account HCI's action on the other electron, 
the calculated spectrum extends even more in 
the transverse direction but new maxima 
are still absent (see figure \ref{figure4}d). 

And only when the action of the HCI's field on the electrons 
is fully included, do new maxima appear 
at $ \vartheta_k \approx 125^0 $ (figure \ref{figure4}a). 
In this case the maxima at $ \vartheta_k = 180^0$ 
further loose in intensity and 
the (relative) extension of the spectrum 
in the direction of large $k_{tr}$ is most pronounced. 

It follows from these results that 
the action of the HCI's field on {\it both} electrons  
is necessary for the appearance of the second maxima 
in the emission spectrum. Therefore, 
the physical origin of these maxima is qualitatively 
different from that of the EET mechanism, which  
proceeds via the interaction in the initial channel 
between the ion and only {\it one} of the atomic electrons.    
However, we could not find a simple physical picture 
for this strong-field mechanism. 
   
The maxima at $\vartheta_k = 180^0$ 
for their appearance need, in principle, neither 
the interaction with the HCI in the initial channel 
nor the interaction between the HCI and the emitted electron. 
According to the results presented in figure \ref{figure4} 
the distortion of the initial two-electron state 
by the field of the HCI and the action of this field 
on the emitted electron just diminish the EEA mechanism. 

Additional information about the transfer-ionization 
can be obtained by calculating cross sections for this process 
using the rest frame of the ion. Let us briefly 
discuss the dependence of the transfer-ionization 
in this frame on the emission angle. 
This can be done by studying the cross section 
$d^2 \sigma/d \varepsilon_p sin\theta_p d\theta_p$,   
where $\varepsilon_p$ and $\theta_p$ are the electron 
emission energy and polar angle in this frame. 

\begin{figure}[t] 
\vspace{-0.45cm} 
\begin{center}
\includegraphics[width=0.55\textwidth]{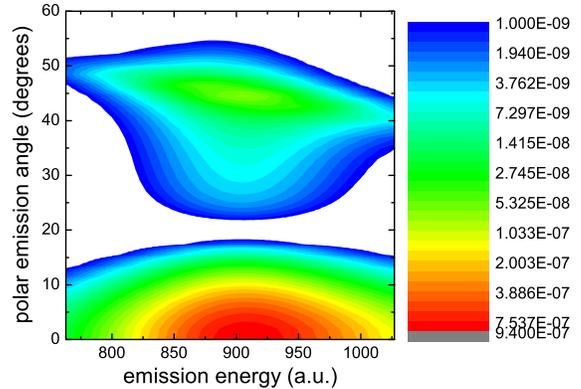}
\end{center}
\vspace{-0.5cm} 
\caption{ \footnotesize{The energy-angular spectrum 
$d^2 \sigma/d \varepsilon_p sin\theta_p d\theta_p$
for the reaction $22.5$ MeV p$^{+}$ + He(1s$^2$) 
$\to$ H(1s) + He$^{2+}$ + e$^-$. 
The spectrum is given in the rest frame of the ion. }}  
\label{figure5} 
\end{figure}

In the rest frame of the ion the EEA and EET mechanisms result 
in the maxima located at 
$ \theta_p \approx 0$ and $ \theta_p \approx 45^0$, respectively  
(see figure \ref{figure5}, \cite{plane-waves}), 
where the angle $ \theta_p$ is counted from the velocity of the atom.

\begin{figure}[t] 
\vspace{-0.45cm}
\begin{center}
\includegraphics[width=0.55\textwidth]{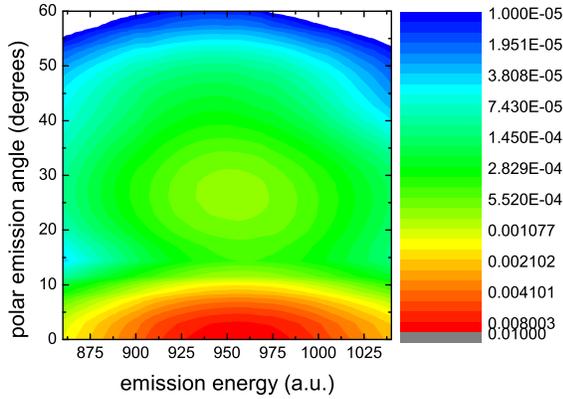}
\end{center}
\vspace{-0.5cm} 
\caption{ \footnotesize{ Same as in figure 
\ref{figure5} but for the reaction $22.5$ MeV/u Ne$^{10+}$ + He(1s$^2$) 
$\to$ Ne$^{9+}$(1s) + He$^{2+}$ + e$^-$. }}  
\label{figure6} 
\end{figure}

\begin{figure}[t] 
\vspace{-0.45cm}
\begin{center}
\includegraphics[width=0.55\textwidth]{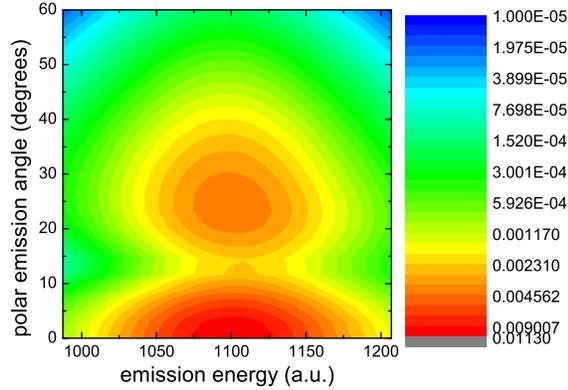}
\end{center}
\vspace{-0.5cm} 
\caption{ \footnotesize{ Same as in figure \ref{figure5} 
but for the reaction $22.5$ MeV/u Ca$^{20+}$ + He(1s$^2$) 
$\to$  Ca$^{19+}$(1s) + He$^{2+}$ + e$^-$. }}  
\label{figure7} 
\end{figure}

\begin{figure}[t] 
\vspace{-0.45cm}
\begin{center}
\includegraphics[width=0.55\textwidth]{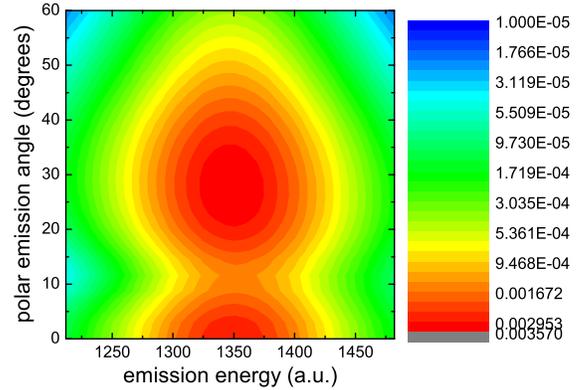}
\end{center}
\vspace{-0.5cm} 
\caption{ \footnotesize{ Same as in figure \ref{figure5} 
but for the reaction $22.5$ MeV/u Zn$^{30+}$ + He(1s$^2$) 
$\to$ Zn$^{29+}$(1s) + He$^{2+}$ + e$^-$. }}  
\label{figure8} 
\end{figure}

In figures \ref{figure6}-\ref{figure8} the energy-angular distribution 
is shown for the correlated transfer-ionization in collisions 
of 22.5 MeV/u Ne$^{10+}$, Ca$^{20+}$ and Zn$^{30+}$ with helium 
resulting in electron capture into the ground state of the ion. 
In all these figures there is one maximum at $ \theta_p \approx 0$ 
corresponding to the EEA, no maximum at $ \theta_p \approx 45^0$ 
and the second maximum whose center is located between 
$ \theta_p \approx 20^0$ and $ \theta_p \approx 30^0$.  
It is seen in the figures that the relative 
intensity of this maximum (compared to the EEA) 
strongly increases when the charge of the ion increases.  

The center of this maximum slightly moves to lower 
angles when we go from Ne$^{10+}$ to Ca$^{20+}$ ions 
but shifts somewhat upwards when Ca$^{20+}$ is replaced by Zn$^{30+}$. 
Thus, for a fixed impact velocity of $v=30$ a.u. the angular position 
of this center depends rather weakly (and nonmonotonously) 
on the charge of the ion when the latter varies 
in the (quite broad) range $10 \leq Z_i \leq 30$.  

In our calculations partial-wave expansions for the states 
$\chi_{{\bf v}_a}({\bf r}_1)$, $\chi_{{\bf v}_a}({\bf r}_2)$ 
and $\chi_{{\bf p}}({\bf r}_2)$  are used \cite{plane-waves-1}. 
For $Z_i \ll v$ the numbers of the partial waves 
become very large. Besides, the radial integrands 
these waves are part of become highly oscillating. 
All this makes it difficult for us 
to perform the computation at $Z_i \ll v$ 
using the Coulomb form for all the states 
$\chi_{{\bf v}_a}({\bf r}_1)$, $\chi_{{\bf v}_a}({\bf r}_2)$ 
and $\chi_{{\bf p}}({\bf r}_2)$.   
Therefore, at the moment we cannot address 
an interesting point: whether with 
decreasing the ratio $Z_i/v$ 
the second maximum smoothly goes over into the EET  
or it decreases in intensity and eventually 
disappears without moving into the region   
$ \theta_p \approx 45^0$ where the EET independently arises. 

\section{ Conclusions } 
       
We have considered transfer-ionization 
in collisions of helium with fast nuclei 
having so high charge $Z_i$ that $Z_i \sim v \gg Z_a$. 
In this consideration we focused on the correlated channels 
of this process in which the electron-electron 
interaction during the collision plays a crucial role. 
Our results were obtained using a treatment which enables one 
to fully account for the action of the strong field 
generated by a highly charged ion on electrons, 
both in the initial and final states of the reaction.   

Our consideration shows that the strong field 
has a very profound effect on the correlated transfer-ionization.  
Compared to the weak-field regime realized 
in collisions with fast low charged ions  
the strong field weakens (in relative terms) 
the emission of high-energy electrons 
into the direction (exactly) opposite to the motion 
of the projectile and strongly suppresses  
the emission perpendicular to this motion. 
Instead, a very substantial part of  
emission via the correlated channels  
goes now into the direction around $\vartheta_k \approx 125^0$ 
and its relative importance 
increases with the strength of the perturbation. 
This is in contrast to collisions with 
fast low charged ions where this part of 
the momentum space is not important at all.  

The correlated transfer-ionization is intimately 
related to the process of radiative two-electron 
transfer in which two atomic electrons are captured by 
the projectile with emission of a single photon. 
Indeed, the electron ejected in the transfer-ionization 
can undergo radiative recombination with the projectile 
leading to the radiative two-electron transfer. 
Since the probability of radiative recombination 
is very low the latter process has much smaller cross sections 
than the transfer-ionization that makes it very difficult 
for experimental observation \cite{winters}.     
Therefore, further theoretical and experimental studies 
of transfer-ionization \cite{experiment} 
may also shed more light on the correlated 
two-electron--one-photon capture.  
  
The author acknowledges support 
from the Extreme Matter Institute EMMI.


\begin{thebibliography}{99}
\bibitem{details} The details of the calculations 
will be given elsewhere.
\bibitem{experiment} According to the numerous 
discussions with my colleagues D. Fischer and P. Mokler  
an experimental verification of 
our theoretical predictions is feasible.
\end{thebibliography}
\end{document}